\documentclass[%
a4paper,
conference,
]{IEEEtran}

\usepackage{import}

\usepackage[utf8]{inputenc}

\usepackage{cite}

\usepackage{ifpdf}

\usepackage{tikz}
\usetikzlibrary{
	decorations.pathreplacing,%
	decorations.markings,%
	positioning,%
	shapes,%
	calc,%
	shapes.arrows,%
	patterns,%
	backgrounds,%
	fit,%
	arrows,%
	automata,%
}

\usetikzlibrary{external}
\ifpdf
	\tikzset{external/system call={pdflatex \tikzexternalcheckshellescape -halt-on-error
			-interaction=batchmode -jobname "\image" "\texsource"}}
	\tikzset{external/up to date check=md5}			%
\else
	\tikzset{external/system call={latex \tikzexternalcheckshellescape -halt-on-error
			-interaction=batchmode -jobname "\image" "\texsource" && 
			dvips -o "\image".eps "\image".dvi}}
	\tikzset{external/up to date check=simple}		%
\fi
\usepackage{marvosym}

\usepackage{fontawesome}
\usepackage{linearb}
\tikzset{radiation/.style={{decorate,decoration={expanding waves,angle=90,segment length=4pt}}}}
\tikzset{antenna/.pic={
		code={
			\draw (0,0) circle (0.1cm);
			\draw[radiation,decoration={angle=60}] (0.1cm,0) -- +(0:0.5);
		}%
}}

\tikzset{font=\sffamily\footnotesize}

\usepackage{pgfplots}
\pgfplotsset{compat=newest}
\usetikzlibrary{positioning, spy,pgfplots.dateplot}

\usepackage{xspace}

\usepackage[binary-units=true]{siunitx}
\usepackage{psfrag}
\usepackage{graphicx}
\usepackage{tabularx,booktabs}
\usepackage{multirow}
\usepackage{rotating}

\usepackage{multirow}

\usepackage[cmex10]{amsmath}

\usepackage{sansmath}

\usepackage[caption=false,font=footnotesize]{subfig}
\usepackage{url}

\usepackage{standalone}		%

\usepackage{import}

\usepackage[
	textsize=tiny,
	]{todonotes}
	\presetkeys{todonotes}{fancyline, color=blue!20}{}
	\tikzset{notestyleraw/.append style={inner sep=2pt, rounded corners=0pt}}
\usepackage[]{acronym}
\usepackage{xspace}

\tikzstyle{background grid}=[draw, black!50,step=1cm]	%

\pgfplotsset{
	box plot/.style={
		/pgfplots/.cd,
		fill=blue!10,
		only marks,
		mark=-,
		mark size=0.15em,
		/pgfplots/error bars/.cd,
		y dir=plus,
		y explicit,
	},
	box plot box/.style={
		/pgfplots/error bars/draw error bar/.code 2 args={%
			\draw  ##1 -- ++(.15em,0pt) |- ##2 -- ++(-.15em,0pt) |- ##1 -- cycle;
		},
		/pgfplots/table/.cd,
		y index=2,
		y error expr={\thisrowno{3}-\thisrowno{2}},
		/pgfplots/box plot
	},
	box plot top whisker/.style={
		/pgfplots/error bars/draw error bar/.code 2 args={%
			\pgfkeysgetvalue{/pgfplots/error bars/error mark}%
			{\pgfplotserrorbarsmark}%
			\pgfkeysgetvalue{/pgfplots/error bars/error mark options}%
			{\pgfplotserrorbarsmarkopts}%
			\path ##1 -- ##2;
		},
		/pgfplots/table/.cd,
		y index=4,
		y error expr={\thisrowno{2}-\thisrowno{4}},
		/pgfplots/box plot
	},
	box plot bottom whisker/.style={
		/pgfplots/error bars/draw error bar/.code 2 args={%
			\pgfkeysgetvalue{/pgfplots/error bars/error mark}%
			{\pgfplotserrorbarsmark}%
			\pgfkeysgetvalue{/pgfplots/error bars/error mark options}%
			{\pgfplotserrorbarsmarkopts}%
			\path ##1 -- ##2;
		},
		/pgfplots/table/.cd,
		y index=5,
		y error expr={\thisrowno{3}-\thisrowno{5}},
		/pgfplots/box plot
	},
	box plot median/.style={
		/pgfplots/box plot
	},
	boxplot/every median/.style={
		ultra thick,dashed,cyan
	}
}

\makeatletter
\def\relativepath{\import@path}
\makeatother

\newcommand{\cf}{cf.\@\xspace}		%
\newcommand{\eg}{e.g.\@\xspace}		%
\newcommand{\ie}{i.e.\@\xspace}		%

\newcommand{\dB}{\decibel}
\newcommand{\dBm}{dBm}

\newcommand{\Fig}[1]{Fig.~#1\xspace}
\newcommand{\Sec}[1]{Sec.~#1\xspace}
\newcommand{\Tab}[1]{Tab.~#1\xspace}

\newcommand{\hex}[1]{\texttt{0x#1}\xspace}

\acrodef{5G}[5G]{fifth generation}
\newcommand{\fiveG}{\ac{5G}\xspace}

\acrodef{4G}[4G]{fourth generation}

\acrodef{mMTC}[mMTC]{massive machine-type communication}

\acrodef{URLLC}[URLLC]{ultra-reliable and low-latency communication}

\acrodef{eMBB}[eMBB]{enhanced mobile broadband}

\acrodef{NWDAF}[NWDAF]{Network Data Analytics Function}
\newcommand{\NWDAF}{\ac{NWDAF}\xspace}

\acrodef{PCF}[PCF]{Policy Control Function}
\newcommand{\PCF}{\ac{PCF}\xspace}

\acrodef{PC}[PC]{Polar Codes}

\acrodef{RAN}[RAN]{Radio Access Network}

\acrodef{SLA}[SLA]{Service Level Agreement}
\newcommand{\SLA}{\ac{SLA}\xspace}

\acrodef{3GPP}[3GPP]{3rd Generation Partnership Project}
\newcommand{\threeGPP}{\ac{3GPP}\xspace}

\acrodef{MNO}{Mobile Network Operator}

\newcommand{\MNOs}{\acp{MNO}\xspace}

\acrodef{ICI}{Inter Cell Interference}
\newcommand{\ICI}{\ac{ICI}\xspace}

\acrodef{RAM}{Random Access Memory}
\newcommand{\RAM}{\ac{RAM}\xspace}

\acrodef{GUI}{Graphical User Interface}
\newcommand{\GUI}{\ac{GUI}\xspace}

\acrodef{CORESET}{Control Resource Set}
\newcommand{\CORESET}{\ac{CORESET}\xspace}
\newcommand{\CORESETs}{\acp{CORESET}\xspace}

\acrodef{C3ACE}[C\textsuperscript{3}\!ACE\xspace]{Client-based Control Channel Analysis for Connectivity Estimation}
\acrodef{EC3ACE}[E-C\textsuperscript{3}\!ACE\xspace]{Enhanced Client-based Control Channel Analysis for Connectivity Estimation}
\newcommand{\CCCACE}{\cccace}
\newcommand{\cccace}{\ac{C3ACE}\xspace}

\acrodef{OWL}[OWL]{Online Watcher for LTE}
\newcommand{\OWL}{\ac{OWL}\xspace}

\newcommand{\LTEye}{LTEye\xspace}

\acrodef{FALCON}[FALCON]{Fast Analysis of LTE Control channels}
\newcommand{\FALCON}{\ac{FALCON}\xspace}

\acrodef{RA}[RA]{Random Access}
\newcommand{\RA}{\ac{RA}\xspace}

\acrodef{RAR}[RAR]{Random Access Response}
\newcommand{\RAR}{\ac{RAR}\xspace}

\acrodef{PDCCH}[PDCCH]{Physical Downlink Control Channel}
\newcommand{\PDCCH}{\ac{PDCCH}\xspace}

\acrodef{EPDCCH}[EPDCCH]{Enhanced \ac{PDCCH}}
\newcommand{\EPDCCH}{\ac{EPDCCH}\xspace}

\acrodef{CPS}[CPS]{Cyber Physical Systems}

\acrodef{WSN}[WSN]{Wireless Sensor Networks}

\acrodef{AP}[AP]{Access Point}

\acrodef{SSB}[SSB]{Swappable Slave Board}

\acrodef{LBT}[LBT]{Listen Before Talk}

\acrodef{SRD}[SRD]{Short Range Devices}

\acrodef{LTE}[LTE]{Long Term Evolution}
\newcommand{\LTE}{\ac{LTE}\xspace}

\acrodef{LTE-A}[LTE-A]{LTE-Advanced}
\newcommand{\LTEA}{\ac{LTE-A}\xspace}

\acrodef{ECDF}[ECDF]{Empirical Cumulative Distribution Function}

\acrodef{RSSI}{Received Signal Strength Indicator}
\acrodef{RSRP}{Reference Signal Received Power}
\acrodef{RSRQ}{Reference Signal Received Quality}
\acrodef{SNR}{Signal to Noise Ratio}
\acrodef{SINR}{Signal to Interference and Noise Ratio}
\acrodef{MCS}[MCS]{Modulation and Coding Scheme}
\newcommand{\MCS}{\ac{MCS}\xspace}
\acrodef{TBS}[TBS]{Transport Block Size}

\acrodef{PRB}[\text{PRB}]{Physical Resource Block}
\acrodef{RB}[\text{RB}]{Resource Block}
\acrodef{CRC}[CRC]{Cyclic Redundancy Check}
\acrodef{DCI}[DCI]{Downlink Control Information}
\acrodef{RNTI}[RNTI]{Radio Network Temporary Identifier}
\acrodef{SI-RNTI}[\ensuremath{\mathrm{SI-RNTI}}]{System-Information \acs{RNTI}}
\acrodef{CCE}[CCE]{Control Channel Element}

\acrodef{UE}[UE]{User Equipment}
\acrodef{eNodeB}[eNodeB]{evolved NodeB}
\acrodef{STG}[STG]{Smart Traffic Generator}
\acrodef{DUT}[DUT]{Device Under Test}
\acrodef{OAI}[OAI]{Open Air Interface}
\acrodef{OFDM}[OFDM]{Orthogonal Frequency Division Multiplexing}
\acrodef{DL}[DL]{Downlink}
\acrodef{UL}[UL]{Uplink}

\acrodef{TCP}[TCP]{Transmission Control Protocol}

\acrodef{UDP}[UDP]{User Datagram Protocol}

\acrodef{FTP}[FTP]{File Transfer Protocol}

\acrodef{CoPoMo}{Context-Aware Power Consumption Model}

\acrodef{BSE}{Base Station Emulator}

\acrodef{CA}{Carrier Aggregation}
\newcommand{\CA}{\ac{CA}\xspace}
\acrodef{CC}{Component Carrier}

\acrodef{MIMO}{Multiple Input Multiple Output}

\acrodef{MU-MIMO}{Multi User-MIMO}
\newcommand{\MUMIMO}{\ac{MU-MIMO}\xspace}
\acrodef{SISO}{Single Input Single Output}

\acrodef{AWGN}{Additional White Gaussian Noise}

\acrodef{COTS}{Commercial Off-the-Shelf}

\acrodef{SDR}{Software-Defined Radio}

\acrodef{IoT}{Internet of Things}

\acrodef{PCC}[PCC]{Primary Carrier Component}

\acrodef{SCC}[SCC]{Secondary Carrier Component}

\newcommand{\RSRP}{\ac{RSRP}\xspace}
\newcommand{\RSRQ}{\ac{RSRQ}\xspace}

\newcommand{\RB}{\ac{RB}\xspace}
\newcommand{\RBs}{\acp{RB}\xspace}

\newcommand{\CRC}{\ac{CRC}\xspace}
\newcommand{\DCI}{\ac{DCI}\xspace}
\newcommand{\RNTI}{\ac{RNTI}\xspace}
\newcommand{\RNTIs}{\acp{RNTI}\xspace}

\newcommand{\CCEs}{\acp{CCE}\xspace}

\newcommand{\UE}{\ac{UE}\xspace}
\newcommand{\UEs}{\acp{UE}\xspace}
\newcommand{\eNB}{\ac{eNodeB}\xspace}
\newcommand{\eNBs}{\acp{eNodeB}\xspace}

\acrodef{RMSE}[RMSE]{Root Mean Square Error}

\tikzexternaldisable

\usepackage[%
	driverfallback=dvipdfm,	%
	hidelinks
	]{hyperref}
\usepackage[IEEE]{copyright-overlay}
\CopyrightAuthor{Robert Falkenberg}
\CopyrightYear{2019}
\ConferenceName{2019 IEEE Global Communications Conference (GLOBECOM)}
\Doi{10.1109/GLOBECOM38437.2019.9014096}
\Bibtex{}
\Bibtex{%
@InProceedings\{Falkenberg2019a,\textCR
	\textHT{}Author = \{Robert Falkenberg and Christian Wietfeld\},\textCR
	\textHT{}Title = \{\{FALCON\}: \{An\} accurate real-time monitor for client-based mobile network data analytics\},\textCR
	\textHT{}Booktitle = \{2019 IEEE Global Communications Conference (GLOBECOM)\},\textCR
	\textHT{}Year = \{2019\},\textCR
	\textHT{}Address = \{Waikoloa, Hawaii, USA\},\textCR
	\textHT{}Month = dec,\textCR
	\textHT{}Doi = \{10.1109/GLOBECOM38437.2019.9014096\},\textCR
	\textHT{}Pages = \{1--7\},\textCR
	\textHT{}Publisher = \{IEEE\}\textCR
	\}
}

\acresetall

\begin{document}

\title{Network Data Analytics Function for Client-based Service Quality Prediction}
\title{Enabling Network Data Analytics Function for Client-based Service Quality Prediction with FALCON}
\title{Network Data Analytics Function for Client-based Service Quality Prediction with FALCON}
\title{Network Data Analytics Function for Client-based Service Quality Prediction with FALCON}
\title{Emulating Client-based Network Data Analytics Function in Public Networks with FALCON}
\title{Performing Client-based Network Data Analytics in Online Networks with FALCON}
\title{Performing Client-based Network Data Analytics in Public Cellular Networks with FALCON}
\title{FALCON: A Method for Client-based Network Data Analytics of Public Cellular Networks}
\title{FALCON: A Method for Network Data Analytics of Public Cellular Networks}
\title{FALCON: An Accurate Real-time Monitor for Client-based Network Data Analytics in Cellular Networks}
\title{FALCON: An Accurate Real-time Monitor for Client-based Mobile Network Data Analytics}

\author{\IEEEauthorblockN{\textbf{Robert Falkenberg and Christian Wietfeld}}
\IEEEauthorblockA{Communication Networks Institute, TU Dortmund University, 44227 Dortmund, Germany\\
Email: \{Robert.Falkenberg, Christian.Wietfeld\}@tu-dortmund.de}%
}

\maketitle

\begin{abstract}

Network data analysis is the fundamental basis for the development of methods to increase service quality in mobile networks.
This requires accurate data of the current load in the network.
The control channel analysis is a way to monitor the resource allocations and the throughput of all active subscribers in a public mobile radio cell.
Previous open-source approaches require either ideal radio conditions or long-term observations in order to obtain reliable data.
Otherwise, the revealed information is polluted by spurious assignments with random content.
In this paper, we present a new open-source instrument for Fast Analysis of LTE Control channels (FALCON),
which combines a  novel shortcut-decoding approach with the most reliable techniques known to us 
to reduce the aforementioned requirements significantly.
Long-term field measurements reveal that FALCON reduces errors in average by three orders of magnitude compared to currently the best approach.
FALCON allows observations at locations with interference and enables mobile applications with single short-term tracking of the local load situation.
It is compatible with numerous software defined radios and can be used on standard computers for a reliable real-time analysis.

\end{abstract}

\IEEEpeerreviewmaketitle

\PrintCopyrightOverlay

\section{Introduction and Related Work}\label{ch:Introduction}

Data analytics, in conjunction with machine learning, is envisioned to empower future mobile networks to predict and avoid congestions through pro-active traffic steering and load balancing.
At the same time, the concept of the network slicing in the \fiveG mobile communication system enables vertical industries to provide tailored services which span over multiple, partially virtualized network components on a shared infrastructure.
With the presence of large sets of heterogeneous service quality requirements in a dynamically changing network, data analytics play a major role not only for network maintenance, but also for avoiding \SLA violations with far-reaching consequences.
For this purpose, the \threeGPP has recently begun to standardize interfaces for a \NWDAF~\cite{3GPP-TS-29.520-NWDA}, which allows network components, \eg the \PCF, to subscribe notifications for events like slice congestions.

In order to not limit the development of these functions to simulated scenarios, researchers also need access to detailed load information of current live networks. 
From this information can be derived realistic load profiles, models for user behavior and the evaluation of prediction methods in the field.
A recent study required knowledge of the cell load for an in-depth analysis of bottlenecks in public \LTE networks~\cite{Bursttracker2019}.
Traces of the cell load were used to evaluate the performance of \CA in \LTEA~\cite{Ludant2017} or to measure the spectral efficiency in the field~\cite{ETRI2018}. 
Artificial neural networks were used in~\cite{Falkenberg2017c} to implement a client-based data rate prediction depending on the current network load.
This information can further improve the efficiency of context-predictive vehicle-to-cloud communication~\cite{Sliwa/etal/2019} and avoid transmissions in congested cells.

\begin{figure}[b!] 
	\vspace{-5pt}
	\centering
	\tikzsetnextfilename{scenario}
	\subimport{../fig/}{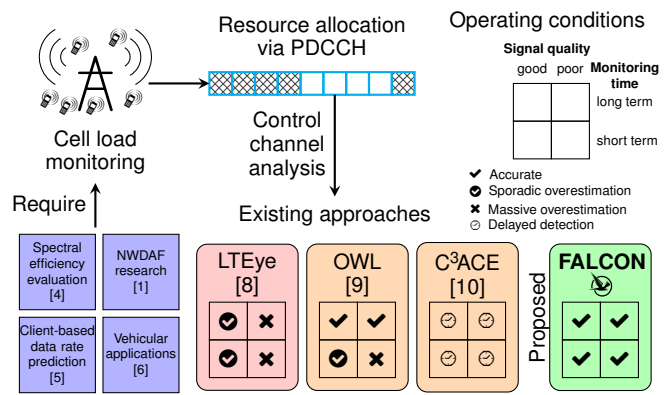}

	\caption{Overview of existing approaches for monitoring the cell load and their suitability for different operating conditions. The proposed method \FALCON enables accurate short-term monitoring even in non-ideal radio conditions.}
	\label{fig:scenario}
\end{figure}%
In order to obtain the necessary information about the cell load, the \PDCCH of an \LTE cell can be analyzed.
The channel signalizes the resource allocation and the \MCS with a resolution of \SI{1}{\milli\second} to individual participants.
Although these messages are not encrypted, only the addressed \UE can verify the integrity of the decoded message because the \CRC sequence is scrambled with the \RNTI of the \UE.
The main difficulty of the control channel analysis is thus the reconstruction of the set of active \RNTIs.

Besides expensive commercial tools with specialized hardware~\cite{ETRI2019} and proprietary software solutions~\cite{AIRSCOPE}, there are also some open-source attempts based on \acp{SDR}. These are \LTEye~\cite{LTEye2014}, \OWL~\cite{OWL2016}, and our previous work for \CCCACE~\cite{Falkenberg2016}, which are all described in more detail in the following section.
\Fig{\ref{fig:scenario}} provides an overview of the key properties of these approaches with regard to operating conditions (signal quality and monitoring time) and lists use cases which require a cell-load monitoring.
\begin{figure}[tbp] 
	\centering		  
	\includegraphics[width=1\columnwidth]{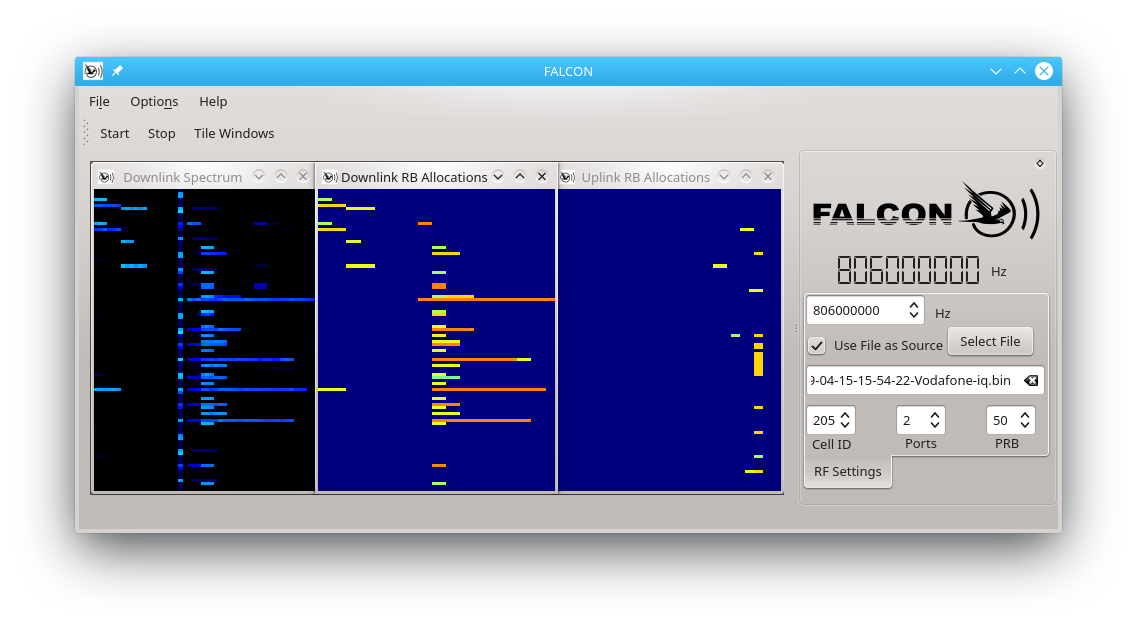}
	\caption{Screenshot of \FALCON's live resource allocation viewer in action. Three waterfall plots show the current signal magnitude of the cell signal (left) and the decoded resource block allocations for the downlink (center) and the uplink (right) at subframe level. Individual users are colored differently. The downlink allocations perfectly match the actual spectral occupation.}
	\label{fig:screenshot}
	\vspace{-5pt}
\end{figure}%

To the best of our knowledge, \OWL is currently the most reliable open-source instrument for continuous long-term moni\-to\-ring, as it tracks the initial \RNTI assignments of new \UE that enter the cell.
To discover \RNTIs from passed assignments, it falls back to \LTEye's re-encoding technique, which is explained in \Sec{\ref{ch:Challenge}}.
However, re-encoding is very sensitive to interference and noise so that \OWL and \LTEye both require almost perfect radio conditions to produce a reliable output in short term observations.
\CCCACE, on the other hand, tracks the activity of supposed \RNTIs with a histogram and only accepts those \RNTIs that exceed a predefined occurrence threshold in the histogram.
While this approach is robust against spurious detections, the time required to exceed the threshold leads to a delayed detection of new \UE and \UE with low activity may stick below the threshold.

In this paper, we present \acs{FALCON}\footnote{The code is available at \url{https://github.com/falkenber9/falcon}}: \acl{FALCON}, an improved, further developed combination of the existing methods, which is suitable for both, long-term and short-term monitoring of \LTE resource allocations in non-ideal radio conditions.
\FALCON is completely open-source and it is released under Affero General Public License v3.

The rest of the paper consists of an introduction in control channel analysis and the involved challenges (\Sec{\ref{ch:CCAnalysis}}), 
an overview of the \FALCON suite with the involved \DCI validation methods (\Sec{\ref{ch:Challenge}}), and a comparative evaluation of \FALCON against \OWL (\Sec{\ref{ch:Evaluation}}). Finally, a conclusion is drawn in \Sec{\ref{ch:Conclusion}}.

\section{Control Channel Analysis}\label{ch:CCAnalysis}
In this section, we give a brief introduction into resource assignment functions in \LTE/\LTEA networks to provide an understanding of the challenges in monitor cell load through observations.
We present the key methods used by existing approaches and discuss the applicability in \fiveG.

\subsection{Resource Assignments}\label{ch:ResourceAssignments}
In \LTE/\LTEA networks, the \eNB centrally governs radio resources in terms of \RBs which are dispensed to the attached \UEs.
Except for the initial network access, namely \RA, the entire communication happens according to explicit resource assignments by the \eNB.
These assignments, namely \DCI, are signaled via \PDCCH in the first symbols of each subframe once in \SI{1}{\milli\second}.
\DCI from \PDCCH in subframe $n$ addresses the current subframe $n$ for the downlink and subframe $n+4$ in uplink direction.
\LTEA extends this signaling by an optional \EPDCCH, which can be located in sets of \RBs distributed along the entire resource grid in order to reduce interference and congestion of the \PDCCH control region in very dense environments or heterogeneous networks.

As shown in \Fig{\ref{fig:recursion}}, the \PDCCH region consists of several \CCEs, which are occupied by the encoded \DCI for distinct \UEs.
Depending on radio conditions of particular \UE, the \eNB selects a suitable aggregation level $L\in \{1,2,4,8\}$ and fills the encoded \DCI into $L$ consecutive \CCEs.
Each \DCI is appended by a \num{16}-bit \CRC checksum, which is additionally scrambled (via bitwise XOR) by the particular \UE identity, \ie the \RNTI, of the same length.
The assignment of \RNTI is performed during the \RA procedure in a not encrypted \RAR message.
Further sets of \RNTIs are defined to serve specific purposes, \eg for system information (SI-\RNTI = \hex{FFFF}) or paging (P-\RNTI = \hex{FFFE}).

Lastly, the standard defines various \DCI formats~\cite{3GPP-TS-36-213-Coding} for allocations in different transmission modes and which differ in their payload size.
The \eNB applies channel coding and rate matching to fit the encoded \DCI exactly into the selected number of \CCEs.

\subsection{Blind Decoding and Search Space}\label{ch:BlindDecoding}
The \PDCCH has no table of contents and neither \DCI format nor aggregation level $L$ is signaled explicitly.
Therefore, the \UE has to repeatedly perform a blind decoding of the \PDCCH contents under the assumption of a certain \DCI format and $L$.
Only after a candidate has been decoded, the \CRC reveals a hit if the \RNTI equals to the \CRC value.

In order to limit the number of decoding attempts, the standard defines two search spaces, which comprise two subsets of all possible locations and associated aggregation levels.
The first search space is \UE-specific and comprises 16 locations which are scattered along the \PDCCH. These depend on the \RNTI and the current subframe number $0\ldots9$ according to the search space function~\cite{3GPP-TS-36.213-PHY}.
The second search space is common for all \RNTI and comprises 6 locations in the first \CCEs of a \PDCCH. Besides \UE-specific data, this search space also carries \DCI for paging, system information and random access.
Ultimately, the \UE only monitors \DCI formats for its current transmission mode.
Carrier Aggregation in \LTEA may lead to additional \UE-specific search spaces for each component carrier in case of cross-carrier scheduling.

\subsection{Approaches and Challenges in DCI Validation}\label{ch:Challenge}

\begin{table}[tbp]
	\centering
	\caption{Overview of DCI Validation Techniques Used by Non-commercial PDCCH Decoders}
	\setlength\extrarowheight{2pt}	
	\begin{tabularx}{\columnwidth}{p{2.0cm}cccc}
		\toprule
		& \multicolumn{4}{c}{\textbf{Decoder}}\\
		\cmidrule(lr){2-5}
		\textbf{Technique} & \textbf{\LTEye}~\cite{LTEye2014} & \textbf{\OWL}~\cite{OWL2016} & \textbf{\CCCACE}~\cite{Falkenberg2016} &  \textbf{\FALCON} \\
		\midrule
		Signal power & X & X & -- & X \\
		Re-encoding & X & X & -- & -- \\
		RAR tracking & -- & X & -- & X \\
		RNTI histograms & -- & -- & X & X \\
		Search space\newline coherence & -- & X\textsuperscript{*} & X & X \\
		Short-cut (\textbf{new}) & -- & -- & -- & X \\
		\bottomrule
		\multicolumn{4}{l}{\textsuperscript{*}Added in a later release}
	\end{tabularx}
	\label{tab:techniques}
	\vspace*{-10pt}
\end{table}

The major challenge of decoding the global set of \DCI in a monitored cell is related to the concept of blind-decoding.
In contrast to a regular \UE, an external observer has no knowledge about currently assigned \RNTIs, individual transmission modes and the associated subset of \DCI formats.
Therefore, rules for search space reduction are not applicable here.
Instead, any possible location needs to be decoded with respect to any potential \DCI format and aggregation level.
This results in a large set of mostly false candidates for the same sequence of \CCEs.
However, since the \CRC is scrambled with the \RNTI of the addressee, a \CRC validation presupposes knowledge of valid \RNTIs.
To overcome these limitations, researchers have proposed different approaches to validate \DCI candidates and reconstruct the set of active \RNTIs:
\begin{LaTeXdescription}
	\item[Signal power detection] greatly reduces the number of blind decoding attempts by disqualifying any \CCEs which undershoot a predefined average signal level. However, \ICI may lead to a number of false-positive classifications especially if the signal is received at the edge between two sectors.
	\item[Re-encoding] the decoded \DCI and comparing the output with the initial bit sequence was proposed with LTEye~\cite{LTEye2014} and is also included as a fall-back in \OWL~\cite{OWL2016}. Candidates are regarded as valid if the encoded sequences match in orders of at least \SI{97}{\percent}. However, our measurements show that this approach is error-prone in the presence of interference, noise or in case of imperfect synchronization which all lead to massive false detections. Therefore, this approach is not included in \FALCON.
	\item[\RAR tracking] was introduced with \OWL~\cite{OWL2016} as a reliable method for long-term observations which captures initial \RNTI assignment messages.
	However, short-term observations lack unseen assignments in the past, since this method discovers only those \RNTIs which have been assigned during the observation period.
	\item[\RNTI histograms] have been presented with \CCCACE~\cite{Falkenberg2016} as a method for \DCI validation in short-term observations, which is less sensitive to the radio conditions. Given a short time window $T$, valid \RNTI appear more frequently in that interval while \RNTI from false candidates are evenly distributed along the 16-bit value range. If an \RNTI exceeds a certain threshold $k$, the corresponding \DCI is assumed as valid. The values $T$ and $k$ are a trade-off between the false-positive probability and the required minimum \UE activity to be detected.
	A threshold value that is set too high filters out subscribers who are very rarely scheduled.
	\item[Search space coherence] can be validated by reverse application of the search space function (\cf \Sec{\ref{ch:BlindDecoding}}) after decoding.
	Since the \eNB never places \DCI outside the associated search space, outlying candidates can be safely discarded.
	This method was introduced with \CCCACE and was added to \OWL afterward.
	With a search space of 22 candidates for a regular \UE (\ie 6 common + 16 \UE-specific) and the maximum number of 88 \CCEs in a \SI{20}{\mega\hertz} cell with one or two antenna ports, this approach filters in average $\sim$\SI{87}{\percent} of all false candidates.
	However, the efficiency of this filter shrinks with the number of \CCEs, \eg down to $\sim$\SI{73}{\percent} for \SI{10}{\mega\hertz} with 44 \CCEs.
	\item[Shortcut decoding] is a novel method of the \FALCON decoder presented in this paper for the rapid validation of \DCI in short-term observations. Uncertain \DCI candidates are decoded once more by using only the first half of their occupied \CCEs. If this results in the same checksum, the \DCI is accepted. The procedure is part of a recursive \DCI search, which is explained in more detail in the next section.
\end{LaTeXdescription}
\Tab{\ref{tab:techniques}} provides a comparison of \FALCON and previous non-commercial decoders with regard to the \DCI validation techniques they contain.

\subsection{Applicability in 5G}\label{ch:5GRelevance}
\fiveG also uses control channels for resource allocation, but \DCI is encoded with polar codes with higher spectral efficiency and lower decoding complexity~\cite{Arikan2009}.
However, a larger search space increases the number of decoding attempts.
The \RNTI still comprises 16~bits, but the checksum of the \DCI has been increased to 24~bits.
After subtracting 3~bits for list decoding of the polar codes, at least 5~bits are not scrambled and can be used for a vague validation without \RNTI knowledge~\cite{Sandberg2018}.
It can be supported by an \RNTI histogram to discover the active set.
However, an additional requirement is the localization of the \CORESETs, which can be located anywhere in the resource grid, analogous to the \EPDCCH.
If the \CORESET makes use of beamforming, a higher receive signal strength may be required to decode the contents correctly.

\section{Structure of FALCON}\label{ch:FALCON} 
In this section, we introduce \acs{FALCON}: \acl{FALCON}.
\FALCON comprises an open source software suite for real-time monitoring of \LTE resource allocations based on \PDCCH decoding  over the air interface.
Besides the decoder and a visualization tool, the software includes a signal recorder with integrated network probing and a remote controller for synchronized capturing of multiple cells or \MNOs.
Additionally, \FALCON includes a port of \OWL's recorder and decoder, the latter of which can also be run in \LTEye mode.

The software is based on the SRSLTE library~\cite{srsLTE2016} v18.12 and
 is kept separated from the underlying library in order to benefit from future updates without a tedious merge and prepares the integration of a 5G library.

FALCON can be executed on an x86-based general purpose computer running a generic Linux kernel.
Any software defined radio supported by SRSLTE can be used to perform over-the-air measurements.
The software is tested with the USRP~B210 by Ettus Research.
Without any radio, the software decoders are also capable to process and visualize recorded signals from a file.

\subsection{Signal Recorder with Network Probing}\label{ch:Recorder}
The signal recorder is an extended version of \OWL's recorder and capable of synchronizing to a particular \LTE cell and capturing the raw I/Q samples for a predefined time interval.
Data is either written directly to a hard disk or buffered into \RAM and written to a hard disk when finished.
Buffering greatly reduces the IO-load on the system and avoids the loss samples from the radio transceiver that would otherwise lead to a loss of cell synchronization.
As buffering consumes \SI{88}{\mega\byte} of RAM per second for a \SI{10}{\mega\hertz} cell, it is applicable for short-term recordings below one or two minutes.

The monitored cell is selected automatically, manually or corresponding to the serving cell of an external modem.
That external modem is also used to produce cell traffic for measurements of the achievable throughput and the involved transmission power.
These results are stored together with cell information and quality indicators such as \RSRP and \RSRQ.
Simultaneous capturing with multiple recorders can be synchronized via Ethernet by an additional remote controller process.

\subsection{Real-Time Decoder}
\begin{figure}[tb]  	
	\centering		  
	\tikzsetnextfilename{recursion}
	\begin{tikzpicture}[font=\sffamily\footnotesize,
					]

\tikzstyle{background grid}=[draw, black!50,step=1cm]

\def\xw{3mm}
\tikzstyle{CCE} = [draw, cyan, inner sep=0, outer sep=0, minimum height=6mm, minimum width=\xw, node distance=0cm, text=black]
\tikzstyle{Cand8} = [draw, cyan, inner sep=0, outer sep=0, minimum height=6mm, minimum width=8*\xw, node distance=0cm, text=black]
\tikzstyle{Cand4} = [Cand8,minimum width=4*\xw]
\tikzstyle{Cand2} = [Cand8,minimum width=2*\xw]
\tikzstyle{Cand1} = [Cand8,minimum width=1*\xw]
\tikzstyle{ggg} = [fill=black!70, pattern=crosshatch,pattern color=black!70]
\tikzstyle{rrr} = [fill=red, pattern=crosshatch dots,pattern color=red]
\tikzstyle{ooo} = [fill=orange!70!white]
\tikzstyle{ddd} = [fill=green!30]

\tikzstyle{label} = [node distance=1mm, inner sep=0, outer sep=0.333em]
\tikzstyle{labelOuter} = [node distance=1mm]
\tikzstyle{bgw} = [fill=white,fill opacity=0.7]%

\node[CCE, ggg] (cce0) {};
\node[CCE, ggg, right=of cce0] (cce1) {};
\node[CCE, ggg, right=of cce1] (cce2) {};
\node[CCE, ggg, right=of cce2] (cce3) {};
\node[CCE, right=of cce3] (cce4) {};
\node[CCE, right=of cce4] (cce5) {};
\node[CCE, right=of cce5] (cce6) {};
\node[CCE, right=of cce6] (cce7) {};
\node[CCE, ggg, right=of cce7] (cce8) {};
\node[CCE, right=of cce8] (cce9) {};
\node[CCE, right=of cce9] (cce10) {};
\node[CCE, right=of cce10] (cce11) {};
\node[CCE, ggg, right=of cce11] (cce12) {};
\node[CCE, ggg, right=of cce12] (cce13) {};
\node[CCE, ggg, right=of cce13] (cce14) {};
\node[CCE, ggg, right=of cce14] (cce15) {};

\node[Cand8, anchor=north west, below=3mm of cce0.south west, anchor=north west] (c80) {1};
\node[Cand8, right=of c80] (c81) {2};

\node[Cand4, rrr, anchor=north west, below=3mm of c80.south west, anchor=north west] (c40) {}; \node[label, bgw, fill opacity=1] at (c40) {3};
\node[Cand4, right=of c40] (c41) {6};
\node[Cand4, right=of c41] (c42) {7};
\node[Cand4, ooo, right=of c42] (c43) {8};

\node[Cand2, ddd, anchor=north west, below=3mm of c40.south west, anchor=north west] (c20) {4};
\node[Cand2, ddd, right=of c20] (c21) {5};
\node[Cand2, right=of c21] (c22) {};
\node[Cand2, right=of c22] (c23) {};
\node[Cand2, right=of c23] (c24) {};
\node[Cand2, right=of c24] (c25) {};
\node[Cand2, ooo, right=of c25] (c26) {9};
\node[Cand2, right=of c26] (c27) {-};

\node[Cand1, anchor=north west, below=3mm of c20.south west, anchor=north west] (c10) {-};
\node[Cand1, right=of c10] (c11) {-};
\node[Cand1, right=of c11] (c12) {-};
\node[Cand1, right=of c12] (c13) {-};
\node[Cand1, right=of c13] (c14) {};
\node[Cand1, right=of c14] (c15) {};
\node[Cand1, right=of c15] (c16) {};
\node[Cand1, right=of c16] (c17) {};
\node[Cand1, rrr, right=of c17] (c18) {}; \node[label, bgw, fill opacity=1, inner sep=0pt, outer sep=0pt] at (c18) {10}; \node[Cand1, right=of c17] (c18overlay) {};
\node[Cand1, right=of c18] (c19) {};
\node[Cand1, right=of c19] (c110) {};
\node[Cand1, right=of c110] (c111) {};
\node[Cand1, right=of c111] (c112) {};
\node[Cand1, right=of c112] (c113) {};
\node[Cand1, right=of c113] (c114) {};
\node[Cand1, right=of c114] (c115) {};

\node[label, left=of cce0] {PDCCH};
\node[label, left=of c80] {$L=8$};
\node[label, left=of c40] {$L=4$};
\node[label, left=of c20] {$L=2$};
\node[label, left=of c10] {$L=1$};

\node[labelOuter, above=of cce0] {CCE\textsubscript{0}};
\node[labelOuter, above=of cce4] {CCE\textsubscript{4}};
\node[labelOuter, above=of cce8] {CCE\textsubscript{8}};
\node[labelOuter, above=of cce12] {CCE\textsubscript{12}};

\node[Cand1, ggg, right= 5mm of cce15] (L0) {} ;
\node[Cand1, below= 1mm of L0] (L1) {} ;
\node[Cand1, rrr, below=1mm of L1] (L2) {};
\node[Cand1, ddd, below= 1mm of L2] (L3) {};
\node[Cand1, ooo, below= 1mm of L3] (L4) {};

\node[label, right=of L0, text width=1.5cm] {Occupied};
\node[label, right=of L1, text width=1.5cm] {Empty or skipped};
\node[label, right=of L2] {Rejected};
\node[label, right=of L3] {Accepted};
\node[label, right=of L4, text width=1.5cm] {Accepted by shortcut};

\draw[-stealth, thick] ([xshift=-0.6mm]c40.west) to node[label, left, align=right] {Inspect\\[-1pt]recursively} ([xshift=-0.6mm]c20.west);
\node[label, bgw, above=0mm of c40, inner sep=0pt, outer sep=0pt, yshift=2pt](tmp) {\phantom{Unknown RNTI}}; \node[label] at (tmp) {Unknown RNTI};	%

\draw[-stealth, thick] ([xshift=0.6mm]c43.east) to ([xshift=0.6mm]c27.east);
\draw[-stealth, thick] (c26.west) to[bend left=70] node[label, bgw, left, align=right,yshift=-2pt] (tmp){\phantom{Match}\\[-1pt]\phantom{parent}\\[-1pt]\phantom{CRC}} (c43.west); \node[label,align=right] at (tmp) {Match\\[-2pt]parent\\[-2pt]CRC};

\node[label, bgw, right=2pt of c18, inner sep=0pt, outer sep=0pt](tmp) {\phantom{Noise}}; \node[label] at (tmp) {Noise};

\end{tikzpicture}

	\caption{Illustration of \FALCON's new recursive \PDCCH decoding procedure. Possible \DCI locations are inspected sequentially starting with the largest aggregation level $L$=$8$. Locations containing empty \CCEs are skipped (1, 2). Checked locations without any known \RNTI (3) are split into two halves and are examined recursively with $L/2$ (4, 5). Locations that overlap any match are also skipped (-). If a shortened location (9) decodes the same uncertain \RNTI as the parent (8), the candidate is accepted immediately.}
	\label{fig:recursion}
	\vspace{-5pt}
\end{figure}
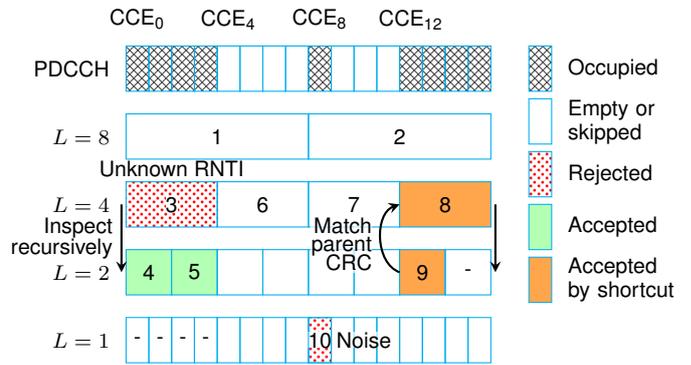

\FALCON's core component is the \PDCCH decoder, which is capable of tracking either an online \LTE signal or an offline recording and reliably decodes any resource assignments from the cell's \PDCCH in real time.
Like a regular \UE, it synchronizes to a cell and configures the receiver according to the cell configuration.
Instead of performing an attach, the decoder remains passive and goes ahead with \PDCCH analysis.
This includes a systematic decoding of any possible location among the \CCEs (including every aggregation level) with all potential \DCI formats.
As stated in \Sec{\ref{ch:Challenge}}, the major challenge is the efficient reconstruction of the valid \RNTI set for a reliable validation of the decoded \DCI candidates.
Similar to \OWL, \FALCON decodes only locations with a sufficient signal power in all covered \CCEs and checks the coherence between the candidate's \RNTI and the corresponding search space.
It also keeps track of any \RAR messages that contain \RNTI assignments for newly joined \UEs which are immediately added to the active set~\cite{OWL2016}.
However, to quickly bootstrap the list of already active \RNTI, \FALCON does not rely on \LTEye's and \OWL's re-encoding technique, that is sensitive to the channel conditions.
Rather, \FALCON combines \RNTI histograms from \CCCACE with the new recursive shortcut validation technique:

Instead of processing all possible locations as breadth-first search with descending aggregation levels like \OWL, \FALCON performs a depth-first search as illustrated in \Fig{\ref{fig:recursion}}.
If a decoded location with aggregation level $L$ does not contain any candidate from the active set (\cf 3), the location is split and inspected recursively using $L/2$ (\cf 4 and 5) until valid candidates are found or $L=1$ is reached.
When the recursion does not detect any valid \DCI, all coherent but rejected \RNTIs along the recursion path are added to a histogram of uncertain \RNTIs.
If an \RNTI appears at least five times within a sliding window of \SI{200}{\milli\second}, it is added to the active set~\cite{Falkenberg2016}.
A shortcut is taken, if decoding a shortened location (\cf 9) results in the same \RNTI as the parent (previously rejected) candidate.
In this case, the \DCI is accepted and the \RNTI is immediately added to the active set.
With this method, the majority of \RNTIs is detected at first occurrence.
This shortcut works because the \eNB implements rate matching of encoded \DCI by circularly writing the encoded sequence into $L$ consecutive \CCEs~\cite{3GPP-TS-36-213-Coding}.

Finally, the decoder writes the content of validated \DCI into a file or visualizes the resource assignments in the subsequent viewer.

\subsection{Real-Time Resource Allocation Viewer}
For a live visualization of the cell activity or for a playback of a previous recording, the \PDCCH decoder is also embedded into an OpenGL-accelerated \GUI as shown in \Fig{\ref{fig:screenshot}}.
Resource allocations for uplink and downlink are displayed at subframe level and can be compared with the allocation in the spectrum.
Additional metrics, such as average throughput or activity of individual users, can also be displayed.

\subsection{Comparative Software and Fairness}\label{ch:fairness}
In order to allow a fair comparison between \FALCON, \OWL and \LTEye\footnote{\LTEye is mimicked by \OWL with disabled \RAR tracking (\cf \Tab{\ref{tab:techniques}})}, \OWL was separated from the outdated library and adapted to the current version of SRSLTE.
To ensure that the port does not break the original functionality, numerous records were analyzed by both the original and the ported version.
Both versions lead to an identical classification of all candidates if the resolution of the Viterbi decoder is equalized.
Only negligible differences in the content of some decoded \DCI were found, which are the result of corrected bugs and which have no influence on the functionality.
The computational complexity of \OWL and \FALCON is almost identical.
Both require approx. \SI{2}{\second} to analyze a record of \SI{5}{\second}.%

\subsection{Privacy}
Although the decoding of \DCI gives the impression that sensitive information is being detected, the privacy of cell users is not at risk.
\RNTIs are volatile identifiers and are released after a few seconds of inactivity.
The payload within the allocated resources is encrypted and cannot be decrypted without knowledge of the secret keys.

\section{Evaluation and Results}\label{ch:Evaluation}
Validating the correct function of a \PDCCH decoder is a major challenge.
A comparison between spectral occupancy and decoded resource allocations does not provide a reliable evaluation metric due to activity in neighboring sectors.
The same applies to the comparison of occupied and decoded \CCEs due to overlapping \acp{PDCCH} of the neighbors.
In addition, the degree of redundancy of the \DCI corresponds to the radio conditions of the addressed subscriber.
Therefore, if the monitored signal is weak, losses of some \DCI must be expected.
However, this should by no means lead to the acceptance of corrupt \DCI with random content.

A reliable metric is the occurrence of collisions due to the erroneous multiple allocations of the same \RBs within a subframe.
\MUMIMO assignments must be excluded from this, but we have not found any assignments with the corresponding \DCI format 1D anyway.

Furthermore, we found in our experiments that the monitored base stations assign new \RNTIs sequentially. 
As a result, active \RNTIs accumulate in a small value range, especially since \RNTIs are released after a short period of inactivity.
The following describes the setup of a long-time measurement, which is then evaluated against the above criteria.

\subsection{Measurement Setup}

\begin{figure}[tb]  	
	\centering		  
	\vspace*{0.01cm}
	\tikzsetnextfilename{mapview}
	\subimport{../fig/}{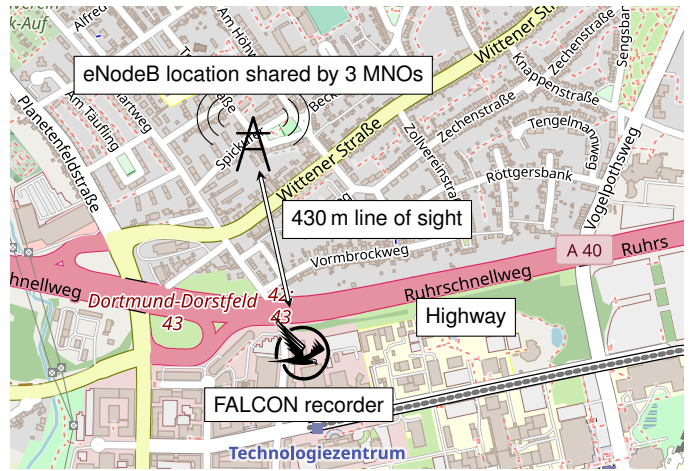}

	\caption{Map of the measurement setup and the three monitored \eNBs. Despite the shared location, the \eNBs differ in transmission power and antenna orientation.
 The cells cover a highway which is densely populated during the rush hours. (Map: \textcopyright OpenStreetMap contributors, CC BY-SA)}
	\label{fig:setup-map}
	\vspace{-5pt}
\end{figure}
For the evaluation of \FALCON, we set up a measurement system with three synchronized recorders to monitor three cells from different \MNOs simultaneously.
Recordings are triggered in intervals of \SI{5}{\minute} to capture \SI{5}{\second} from each cell %
and are directly processed by the decoders \FALCON and \OWL.
We removed recordings that suffered a synchronization loss. %

The recorders were placed in an office next to an insulating window in line of sight to a building which is populated by \eNBs from all three \MNOs (\cf \Fig{\ref{fig:setup-map}}). %
Although the location is identical, the \eNBs differ in antenna orientation and transmit power.
This results in different received signal levels at the measurement point as listed in \Tab{\ref{tab:parameters}}.
The line of sight is crossed by a highway which leads to the German city Dortmund and is intensively used by commuters during the rush hours.
The measurement covers a period of five days, including a weekend, a public holiday and a working day.

\begin{table}[tbp]
	\vspace*{-5pt}
	\centering
	\caption{Key network parameters of the long-term measurement.}
	\setlength\extrarowheight{2pt}	
	\begin{tabularx}{\columnwidth}{p{2.5cm}XXX}
		\toprule
		
		 & \textbf{MNO1} & \textbf{MNO2} & \textbf{MNO3} \\
		\midrule
		Signal Quality & Good & Fair & Poor \\
		\midrule
		Frequency & \SI{800}{\mega\hertz} & \SI{800}{\mega\hertz} & \SI{800}{\mega\hertz}\\
		Bandwidth & \SI{10}{\mega\hertz} & \SI{10}{\mega\hertz} & \SI{10}{\mega\hertz}\\
		\RSRP (average) & \SI{-91.23}{\dBm} & \SI{-99.23}{\dBm} & \SI{-107.69}{\dBm} \\
		\RSRQ (average) & \SI{-7.11}{\dB} & \SI{-10.05}{\dB} & \SI{-14.39}{\dB} \\
		\cmidrule{1-4}
		Measurements & 1389 & 1227 & 1150 \\

		\bottomrule
	\end{tabularx}
	\label{tab:parameters}
	\vspace*{-5pt}
\end{table}

\begin{table}[tbp]
	\vspace*{-5pt}
	\centering
	\caption{Average fraction of subframes with contradictory resource allocation because of false detections. Figures in percent.}
	\setlength\extrarowheight{2pt}	
	\begin{tabularx}{\columnwidth}{p{2.5cm}XXX}
		\toprule
		
		 & \textbf{MNO1} & \textbf{MNO2} & \textbf{MNO3} \\
		\midrule
		Signal Quality & Good & Fair & Poor\\
		\midrule
		\textbf{Uplink} & & & \\
		\cmidrule{1-1}
		\OWL & \SI{0.002}{\percent}  & \SI{0.001}{\percent} & \SI{0.024}{\percent}\\
		\FALCON & \SI{0.000}{\percent} & \SI{0.000}{\percent} & \SI{0.001}{\percent} \\
		\cmidrule{1-4}
		\textbf{Downlink} & & & \\
		\cmidrule{1-1}
		\OWL & \SI{0.284}{\percent} & \SI{0.516}{\percent}& \textbf{\SI{2.527}{\percent}} \\
		\FALCON & \SI{0.000}{\percent}  & \SI{0.001}{\percent} & \textbf{\SI{0.005}{\percent}}\\
		
		\bottomrule
	\end{tabularx}
	\label{tab:collisions}
\end{table}

\subsection{Reliability and False Detections}
Our measurements showed that especially poor radio conditions provoke false \DCI detections which lead to conflicting allocations %
of the same resources to multiple \RNTIs in the same subframe.
\Tab{\ref{tab:collisions}} shows the average fraction of subframes that contained such a collision in uplink and downlink direction.
The results show, that in case of poor radio conditions (MNO3) \FALCON outperforms \OWL in average by three orders of magnitude.
Compared to the downlink, the probability of uplink collisions is smaller because uplink allocations are bound to a single \DCI format while the remaining \DCI formats carry downlink allocations.
Since not every spurious \DCI causes an actual collision, the collisions only indicate a lower bound for false detections.

Therefore, we inspect the number of occurrences of each \RNTI in the decoded time interval in more detail.
The underlying assumption is that spurious \DCI, which is mistakenly assumed as valid, contains a random payload and a random \CRC that is interpreted as \RNTI.
These spurious \RNTIs are scattered uniformly along the entire value range of $2^{16}$ with a very low frequency, each.
The low frequency is a consequence of the small probability hitting the same \RNTI multiple times across a number of coin toss experiments~\cite{Falkenberg2016}.

\begin{figure}[!tb]  	
	\vspace*{0.02cm}
	\centering		  
	\tikzsetnextfilename{RNTI-frequency-poor-rush-stacked}
	\begin{tikzpicture}[
font=\sffamily\footnotesize,
]
\begin{sansmath}

\tikzstyle{label} = [draw, fill=white]

\def\xZ{9500}
\def\xa{13000}
\def\xb{16500}
\def\xc{20000}
\def\xd{28000}

\def\chartDist{5.5cm}

\begin{axis}[
	name=ax1,
	width= 252.0pt,			%
	height=6cm,
	xmin=1, xmax=65536,
	ymax=6e3,
	ymode=log,
	xtick={0,16384,32768,49152,65536},
	xticklabels={0x0,0x4000,0x8000,0xC000,0xFFFF},
	scaled x ticks=false,		%
	xlabel={RNTI},
	ylabel={Frequency in \SI{5}{\second} window},
	ylabel near ticks,
	grid=both,
	legend cell align=left,
	legend pos=north east,
	]

	\addplot[color=red, only marks, mark=x] plot table[col sep=comma, x = rnti, y = frequency] {\relativepath/RNTI-frequency/RNTI-Frequency-2019-04-23-08-04-40-Vodafone-rsrp-108.60-rsrq-13.70-owl.csv};
	\addlegendentry{OWL};
	
	\addplot[color=blue, only marks, mark=o] plot table[col sep=comma, x = rnti, y = frequency] {\relativepath/RNTI-frequency/RNTI-Frequency-2019-04-23-08-04-40-Vodafone-rsrp-108.60-rsrq-13.70-falcon.csv};
	\addlegendentry{FALCON};

	\draw[draw=black, thick] (axis cs:\xZ,0.5) coordinate (boxAul) rectangle ++(axis direction cs:3500, 9e3) coordinate (boxAor);

	\draw[stealth-] (axis cs:\xa, 1e2) to node[label, at end, text width=2cm, anchor=west] {Active RNTI concentrate in tight interval} +(1cm, 0cm);
	
	\node[label, align=center] (false) at (axis cs:55000, 5e1) {OWL false\\dectections};
	\node[below=1mm of false] {$\cdots$};
	\draw[-stealth] (false) to +(-1cm, -1cm);
	\draw[-stealth] (false) to +(-0.4cm, -0.7cm);
	\draw[-stealth] (false) to +(0.4cm, -0.7cm);

\end{axis}

\begin{axis}[
	yshift=-1*\chartDist,
	name=ax2,
	width= 252.0pt,			%
	height=6cm,
	xmin=1, xmax=65536,
	ymax=6e3,
	ymode=log,
	xtick={0,16384,32768,49152,65536},
	xticklabels={0x0,0x4000,0x8000,0xC000,0xFFFF},
	scaled x ticks=false,		%
	xlabel={RNTI},
	ylabel={Frequency in \SI{5}{\second} window},
	ylabel near ticks,
	grid=both,
	legend cell align=left,
	legend pos=north east,
	]

	\addplot[color=red, only marks, mark=x] plot table[col sep=comma, x = rnti, y = frequency] {\relativepath/RNTI-frequency/RNTI-Frequency-2019-04-23-08-09-40-Vodafone-rsrp-106.90-rsrq-13.90-owl.csv};
	\addlegendentry{OWL};
	
	\addplot[color=blue, only marks, mark=o] plot table[col sep=comma, x = rnti, y = frequency] {\relativepath/RNTI-frequency/RNTI-Frequency-2019-04-23-08-09-40-Vodafone-rsrp-106.90-rsrq-13.90-falcon.csv};
	\addlegendentry{FALCON};
	
	\draw[draw=black, dashed] (axis cs:\xZ,0.5) coordinate (boxAAul) rectangle ++(axis direction cs:3500, 9e3) coordinate (boxAAor);
	\draw[draw=black] (axis cs:\xa,0.5) coordinate (boxBBul) rectangle ++(axis direction cs:3500, 9e3) coordinate (boxBBor);

	\draw[-stealth, ultra thick, blue] (axis cs:10800,7e0) to node[above,at end,xshift=1.1cm] {Moving peak over time} +(1.5cm, 0cm);

\end{axis}

\begin{axis}[
	yshift=-2*\chartDist,
	name=ax3,
	width= 252.0pt,			%
	height=6cm,
	xmin=1, xmax=65536,
	ymax=6e3,
	ymode=log,
	xtick={0,16384,32768,49152,65536},
	xticklabels={0x0,0x4000,0x8000,0xC000,0xFFFF},
	scaled x ticks=false,		%
	xlabel={RNTI},
	ylabel={Frequency in \SI{5}{\second} window},
	ylabel near ticks,
	grid=both,
	legend cell align=left,
	legend pos=north east,
	]

	\addplot[color=red, only marks, mark=x] plot table[col sep=comma, x = rnti, y = frequency] {\relativepath/RNTI-frequency/RNTI-Frequency-2019-04-23-08-14-40-Vodafone-rsrp-107.40-rsrq-12.80-owl.csv};
	\addlegendentry{OWL};
	
	\addplot[color=blue, only marks, mark=o] plot table[col sep=comma, x = rnti, y = frequency] {\relativepath/RNTI-frequency/RNTI-Frequency-2019-04-23-08-14-40-Vodafone-rsrp-107.40-rsrq-12.80-falcon.csv};
	\addlegendentry{FALCON};
	
	\draw[draw=black, dotted] (axis cs:\xZ,0.5) coordinate (boxAAAul) rectangle ++(axis direction cs:3500, 9e3) coordinate (boxAAAor);
	\draw[draw=black, dashed] (axis cs:\xa,0.5) coordinate (boxBBBul) rectangle ++(axis direction cs:3500, 9e3) coordinate (boxBBBor);
	\draw[draw=black] (axis cs:\xb,0.5) coordinate (boxCCCul) rectangle ++(axis direction cs:3500, 9e3) coordinate (boxCCCor);
	
	\draw[-stealth, ultra thick, blue] (axis cs:10800,7e0) to node[above,at end,xshift=1.1cm] {Moving peak over time} +(1.5cm, 0cm);

\end{axis}

\draw[-stealth, bend right, blue, thick] ([xshift=-5mm]ax1.south west) to node[right] {\textbf{5min later}} ([xshift=-5mm]ax2.north west);
\draw[-stealth, bend right, blue, thick] ([xshift=-5mm]ax2.south west) to node[right] {\textbf{5min later}} ([xshift=-5mm]ax3.north west);

\draw[dotted] (boxAul) -- (boxAAul |- boxAAor);
\draw[dotted] (boxAul -| boxAor) -- (boxAAor);

\draw[dotted] (boxAAul) -- (boxAAAul |- boxAAAor);
\draw[dotted] (boxAAul -| boxAAor) -- (boxAAAor);
\draw[dotted] (boxBBul) -- (boxBBBul |- boxBBBor);
\draw[dotted] (boxBBul -| boxBBor) -- (boxBBBor);

\end{sansmath}
\end{tikzpicture}

	\vspace{-15pt}
	\caption{Distribution and Frequency of \RNTIs from detected \DCI by \OWL and \FALCON for three subsequent recordings over \SI{5}{\second} in intervals of \SI{5}{\minute}. The recordings were made at poor radio conditions (MNO3, \RSRP: \SI{109}{\dBm}, \RSRQ: \SI{-14}{\dB}) during rush hour. True \RNTIs concentrate in a dense peak region that moves rightwards over time. This indicates a sequential \RNTI assignment by the \eNB and a high user fluctuation in the cell. In contrast to \FALCON, \OWL detects numerous false \RNTIs which form an evenly distributed noise floor.}
	\label{fig:eval:rnti-frequency-poor}
	\vspace{-15pt}
\end{figure}
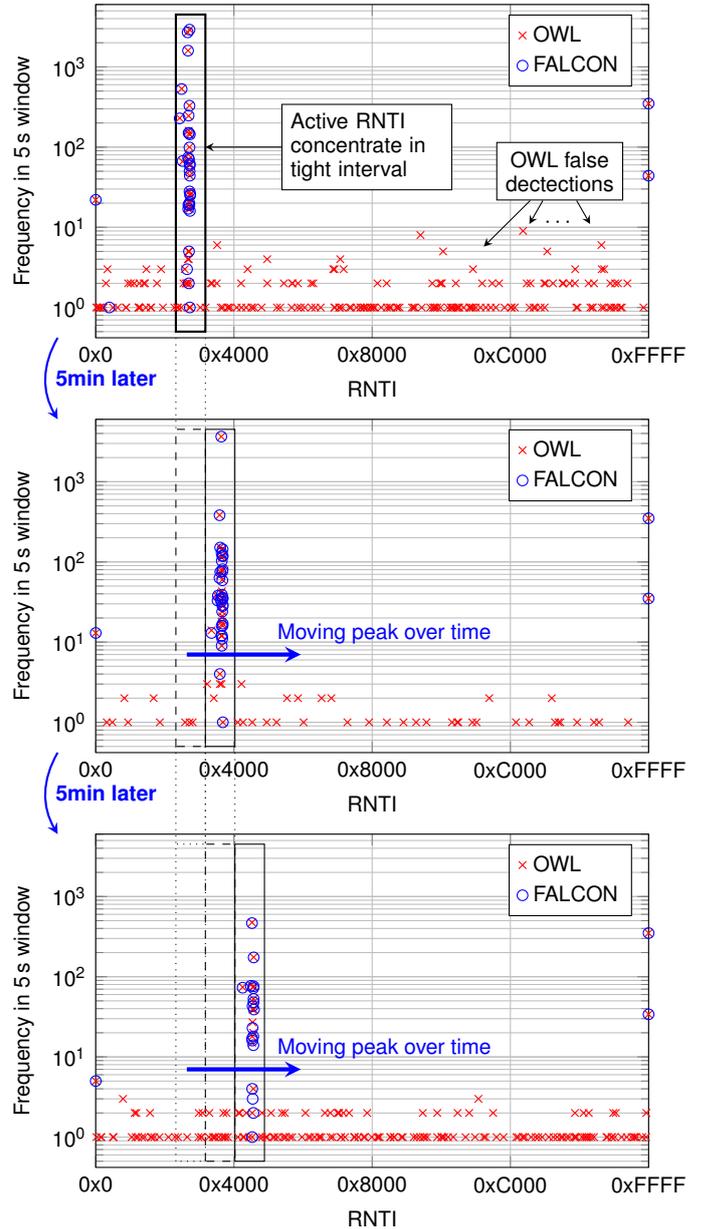

Based on the decoded \DCI by both decoders, we computed the set of \RNTIs which appeared during a monitored interval of \SI{5}{\second} and counted their occurrences.
\Fig{\ref{fig:eval:rnti-frequency-poor}} shows the frequency and the distribution of discovered \RNTIs for three consecutive recordings in the morning rush hour of a working day for the case of poor radio conditions.
In particular, the probing modem (\cf \Sec{\ref{ch:Recorder}}) reported an \RSRP of \SI{109}{\dBm} and an \RSRQ of \SI{-14}{\dB} for this example.
We can see in each chart that the most active \RNTIs concentrate in a small and dense range of values. %
In contrast to \FALCON, \OWL additionally reports a huge set of \RNTIs with a very small frequency of mostly less than three occurrences.
These are evenly distributed across the full value range and match the previous assumptions of random \CRC{}s.
Hence, these \RNTIs are likely false detections from the re-encoding approach.

The peak region, however, is not bound to a fixed interval but moves with time in ascending order along the value range.
To illustrate the progression, we added help lines to \Fig{\ref{fig:eval:rnti-frequency-poor}} to highlight the peak regions from recordings in the same cell five and ten minutes earlier.
This indicates that the base station incrementally assigns \RNTIs to new \UE that join the cell via handover or wake up from idle mode.
In the given example the \RNTI number has advanced by \num{7552} in \num{10} minutes
which represents a mean \RNTI assignment rate of \num{12.6} \RNTIs per second.
Furthermore, each assignment involves an \RA procedure in one of the serving sectors of the cell.
Hence, the \RNTI progression rate must correspond to the monitored number of activities by RA-\RNTIs.
The \RA activity in the \SI{10}{\minute} interval is in average \num{2.66} assignments per second.
Assuming an equal activity in all six cell sectors (three in band 20 and three in band 7), the extrapolated value of $6\cdot2.66=15.96$ slightly overshoots the \RNTI progression rate.
However, only four sectors are directed towards the populated highway, while the remaining sectors cover a residential area.
Considering only those four sectors for the extrapolation, we receive a rate of \num{10.64} assignments per second, which slightly undershoots the progression rate.
Consequently, the two estimations tightly enclose the \RNTI progression rate and confirm the sequential \RNTI assignment by the base station.

Finally, the peak region of active \RNTIs sharply dies away towards smaller (and older) values for both decoders.
This indicates a very short activity time of the \UEs for the purpose of a data exchange followed by a handover into the next cell along the highway or releasing the \RNTI due to inactivity.
Furthermore, the distribution and frequency of the \RNTI noise using \OWL have an equal shape on both sides of the peak region with no significant cumulation on the left side.
This indicates,  that \FALCON does not filter out any meaningful \DCI for \RNTI in the value range beyond the peak region.

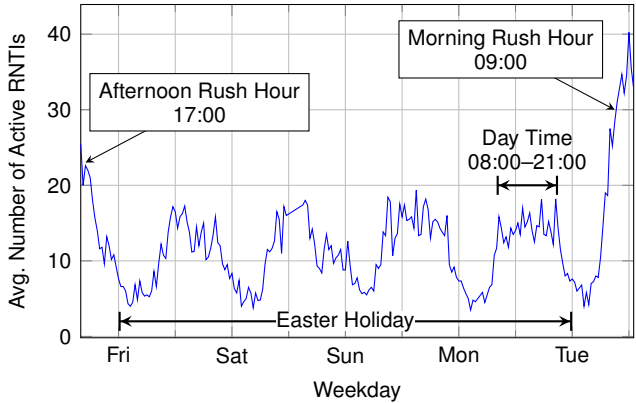
\begin{figure}[!tb]  	
	\vspace*{0.02cm}
	\centering		  
	\tikzsetnextfilename{Number-of-UEs}
	\newcount\tmpCnt

\begin{tikzpicture}[
font=\sffamily\footnotesize,
]
\begin{sansmath}

\begin{axis}[
	width= 252.0pt,			%
	height=6cm,
	date coordinates in=x,
	date ZERO=2019-04-18,
	xmin=2019-04-18 16:00:00,
	xmax=2019-04-23 13:00:00,
	xticklabel          = \pgfcalendardatetojulian{\year-\month-\day}{\tmpCnt}\pgfcalendarjuliantoweekday{\tmpCnt}{\tmpCnt}\pgfcalendarweekdayshortname{\tmpCnt},
	scaled x ticks=false,		%
	minor x tick num=1,
	xlabel={Weekday},
	ylabel={Avg. Number of Active RNTIs},
	ylabel near ticks,
	grid=both,
	]

	\tikzstyle{label} = [draw, fill=white]
	
	\addplot[color=blue] plot table[col sep=comma, x = date, y = tot_nof_UE_dl] {\relativepath/CellLoad/TrainingData-falcon-dci.csv-O2-DE-tot_nof_UE_dl-DL.csv};

\draw[|stealth-{stealth}|,thick] (axis cs: 2019-04-19 00:00:00,2) to node[fill=white,inner sep=0pt] {Easter Holiday}  (axis cs: 2019-04-23 00:00:00,2);
\draw[|stealth-{stealth}|,thick] (axis cs: 2019-04-22 08:00:00,20) to node[inner sep=0pt,above=0.2cm,align=center] {Day Time\\08:00--21:00}  (axis cs: 2019-04-22 21:00:00,20);
\draw[stealth-] (axis cs:2019-04-23 09:00:00,30) to +(-1.5cm,0.8cm) node[label, align=center] {Morning Rush Hour\\09:00};
\draw[stealth-] (axis cs:2019-04-18 17:00:00,23) to +(1.5cm,0.8cm) node[label, align=center] {Afternoon Rush Hour\\17:00};

\end{axis}

\end{sansmath}
\end{tikzpicture}

	\vspace*{-10pt}
	\caption{Application example of \FALCON showing the average number of concurrently active subscribers in a cell of MNO2 during the Easter Weekend. The graph shows a typical slope of $\sim$15 concurrent \UEs during day time on a holiday and more than 30 \UEs during rush hours on a business day.}
	\label{fig:eval:number-of-users}
	\vspace{-15pt}
\end{figure}

\subsection{Application Example}\label{ch:application-example}
An application example of \FALCON is given in \Fig{\ref{fig:eval:number-of-users}} which shows the average number of simultaneously active \UEs in a cell of MNO2 during our campaign.
We define the number of active \UEs as the cardinality of the set of \RNTIs which are scheduled at least once in the monitoring interval of \SI{5}{\second}. Due to a large number of samples, we averaged the results in bins of \SI{30}{\minute}.
The chart shows approx. 15 concurrent \UEs during day time on a holiday and approx. 5 \UEs during night.
Business days, especially during rush hours, are characterized by a significantly higher number of concurrent \UEs.

\section{Conclusion}\label{ch:Conclusion}
In this work, we presented \FALCON, an open-source instrument for  \acl{FALCON} in public \LTE networks.
\FALCON reliably obtains the entire resource allocations of a monitored cell in real-time.
The novel approach of shortcut-decoding provides a fast integrity check of \DCI addressing previously unseen \RNTIs which almost instantly reconstructs the list of currently active \RNTIs in the cell.
This method fully replaces the re-encoding validation technique used by \LTEye~\cite{LTEye2014} and \OWL~\cite{OWL2016}, that is responsible for numerous false detections under less than ideal radio conditions.
With the support of the histogram-based validation of uncertain \DCI candidates, \FALCON maintains its accuracy even in case of a weak signal.
According to our measurements, \FALCON reduces the probability of downlink \RB collisions by three orders of magnitude in case of a poor signal.
Therefore, \FALCON paves the way for reliable short-term monitoring, \eg to supply mobile vehicles with information about cell congestions and increase the prediction accuracy for opportunistic vehicle-to-cloud transmissions.
It is a powerful solution for the analysis of network congestions and a key enabler for the development of \NWDAF.
It allows other researchers to derive realistic traffic models from public networks, independent from the network operators, and without the requirement of expensive hard- and software.

\section*{Acknowledgment}
\footnotesize
Part of the work on this paper has been supported by Deutsche Forschungsgemeinschaft
(DFG) within the Collaborative Research Center SFB 876 ``Providing Information by Resource-Constrained
Analysis'', project A4.

\bibliographystyle{IEEEtran}
\bibliography{manuscript}

\end{document}